# Comment on "Universal decoherence due to gravitational time dilation"

Yuri Bonder,[1] Elias Okon,[2,*] and Daniel Sudarsky[1]
[1]Instituto de Ciencias Nucleares, Universidad Nacional Autónoma de México
Apartado Postal 70-543, México D.F. 04510, México
[2]Instituto de Investigaciones Filosóficas, Universidad Nacional Autónoma de México
Circuito Maestro Mario de la Cueva s/n, México D.F. 04510, México

The interface between gravitation and quantum theory is a fascinating subject. However, it is also riddled with subtleties and slight confusion can easily lead to questionable conclusions. A dramatic example in this regard is provided by [1] where it is claimed that gravitational effects generically produce a novel form of decoherence for systems with internal degrees of freedom, which would account for the emergence of classicality. The effect is supposed to arise from the different gravitational redshifts suffered by such systems when placed in superpositions of positions along the direction of the gravitational field. There are, however, serious issues with the arguments of the paper.

To begin with, the results of [1] cannot be right in light of the *equivalence principle*, which is valid, by construction, in the frameworks employed. This is because the only external force acting on all studied systems is that of a gravitational field with no relevant space-time curvature. As a result, the situations of interest can be analyzed in a free falling frame, in which the systems under study are gravity-free and isolated. Clearly, such scenarios cannot lead to decoherence, as, without gravity, there is nothing to cause it. One might claim that all this shows is that decoherence is frame-dependent. Note however that in order to claim that decoherence has occurred, it is not enough to show that, in the description of the situation according to certain observers, the reduced density matrix for the center of mass is almost diagonal. What one would need to show is that interference is not present in a concrete experiment, a fact that is clearly frame-independent, and not at all shown in [1]. Moreover, since the systems described in [1] are subject to gravity, they will not remain static when placed in a superposition of fixed positions. Of course, one could achieve this by including a compensating force generated by an external device, and this additional interaction may lead to decoherence, but this effect cannot be ascribed to gravity.

Next, notice that the results presented in [1] crucially depend on considering systems where the various internal energy levels contribute to their effective mass, which thus can have more than one value. However, as shown in [2], the nonrelativistic treatment of such situations can lead to spurious relative phases, calling for extreme care in dealing with nonrelativistic approximations. Such a matter is completely ignored by the free use of nonrelativistic quantum mechanics in [1].

Finally, the widespread believe that decoherence can explain the quantum-to-classical transition, which is key in the analysis of the paper, is unjustified [3]. The confusion arises from the fact that the density matrix of an *improper* mixture (which represents the partial description of a subsystem that is part of a larger system in a pure state) has, after decoherence takes place, the same form as that of a *proper* mixture (which represents an actual ensemble of systems) [4]. However, from such formal similarity it does not follow that the two *physical* situations are identical. Therefore, even if the reduced density matrix for the center of mass of the systems considered in [1] has the same form as a statistical mixture, it does not follow that their physical situation is indistinguishable from that of an ensemble; the center of mass continues to be as delocalized as it was before the alleged decoherence takes place.

We conclude from all this that the claims of [1], regarding a gravity induced emergence of classicality, are invalid.

*eokon@filosoficas.unam.mx